\documentclass[preprint,12pt]{elsarticle}

\usepackage{amsmath,amssymb,amsfonts}
\usepackage{graphicx}
\usepackage{booktabs}
\usepackage{multirow}
\usepackage{siunitx}
\usepackage{algorithm}
\usepackage{algpseudocode}
\usepackage{float}
\usepackage[section]{placeins}
\usepackage{url}
\usepackage{hyperref}
\usepackage{enumitem}
\usepackage{mathtools}
\usepackage{tikz}
\usetikzlibrary{arrows.meta,positioning,shapes.geometric,fit,calc}
\journal{Applied Energy}

\begin{document}

\begin{frontmatter}


\title{Feasibility-Aware  Energy Management of a Hydrogen-Enabled Community Microgrid: A Proof-of-Concept Study}

\author[inst1]{Mohamed Atef}
\ead{Muhamed.Atef1408@gmail.com}

\author[inst1]{Sanath Alahakoon}
\author[inst2]{Umme Mumtahina}
\author[inst2]{Peter Wolfs}
\author[inst3]{Tamer Khatib}
\author[inst4]{Moslem Uddin}

\address[inst1]{School of Engineering and Technology, Central Queensland University, Gladstone, QLD 4680, Australia}
\address[inst2]{School of Engineering and Technology, Central Queensland University, Rockhampton, QLD 4680, Australia}
\address[inst3]{Energy Engineering \& Environment Dept., An-Najah National University, Nablus, West Bank 97300, Palestine}
\address[inst4]{School of Engineering \& Technology, The University of New South Wales, Canberra, ACT 2610, Australia}

\begin{abstract}

Hydrogen-enabled community microgrids require coordinated control of intermittent generation and coupled battery--hydrogen storage. This paper presents a feasibility-aware proximal policy optimization (PPO) energy management system for a grid-connected microgrid with photovoltaic and wind generation, battery storage, an electrolyzer, a hydrogen tank, a fuel cell, and diesel backup. Raw continuous actions are projected onto the feasible operating set before the hourly power balance is evaluated. The proof-of-concept study uses 8,760 observations for a 1,000-household community in Rockhampton, Australia; the same chronology and one random seed were used for training and evaluation. In the baseline reliability case with 1\% independent hourly outage probability, the annual net operating cash balance was A\$195,690.67, load satisfaction was 99.77\%, and the gross renewable share of primary supply was 91.2\%. After removing duplicated hydrogen-electricity emissions, the inventory was 1.342~kt~CO$_2$, equivalent to 0.328~kg~CO$_2$/kWh of served community demand and 0.087~kg~CO$_2$/kWh of export-inclusive delivered energy. Increasing hourly outage probability to 5\% reduced the cash balance to A\$169,892.21 and load satisfaction to 98.79\%; battery discharge and diesel generation increased substantially more than fuel-cell output. The results document feasible dispatch for the studied chronology but require unseen testing, multiple seeds, controller benchmarks, export limits, and sustained-outage experiments for broader validation.

\end{abstract}


\end{frontmatter}

\section{Introduction}

\subsection{Background}
Microgrids have emerged as an important architecture for decentralizing electricity supply, improving local reliability, and supporting the transition from conventional centralized grids toward low-carbon distributed energy systems \cite{uddin2023microgrids,chartier2022microgrid,trivedi2022community}. In residential and community applications, microgrids coordinate distributed energy resources (DERs), energy storage systems (ESSs), controllable loads, and grid exchange so that local renewable generation can be used more effectively while maintaining secure supply \cite{atef2026energy,Singh2024}. Recent sectoral reviews indicate that residential microgrids are no longer limited to small PV--battery installations; instead, they are evolving into multi-energy platforms that integrate solar PV, wind generation, battery energy storage systems (BESSs), electric vehicles, demand response, and hydrogen-based storage pathways \cite{atef2026energy,atef2026review,sadeghian2024energy}.

Battery storage is widely used to smooth short-term mismatches between renewable generation and demand, support peak shaving, and improve self-consumption. However, battery-only solutions may be insufficient for prolonged renewable shortages or seasonal energy shifting. Hydrogen technologies therefore provide a complementary long-duration storage pathway, where surplus renewable electricity can be converted through electrolyzers, stored as hydrogen, and later reconverted to electricity through fuel cells \cite{modu2023systematic,elkhatib2024green}. Recent studies further show that hydrogen-based storage can support sustainable and techno-economic operation of renewable-rich systems when its conversion losses and storage constraints are properly managed \cite{das2024sustainable,enaloui2025techno}. This hybrid battery--hydrogen structure is particularly relevant for community microgrids because batteries can handle fast operational balancing, while hydrogen storage can improve autonomy and resilience over longer time scales.

The need for such coordinated storage is reinforced by the uncertainty inherent in renewable-rich community systems. Solar irradiance and wind speed fluctuate continuously, household demand varies with occupant behavior, and electricity tariffs or grid conditions may change across operating intervals. These factors complicate real-time supply--demand balancing and make fixed dispatch rules less effective under dynamic conditions \cite{nguyen2022predictive,uddin2024storage,Witharama2024}. The related Australian residential studies further show that hybrid configurations combining PV, wind, BESS, grid support, backup generation, and hydrogen technologies can reduce cost and emissions, but their benefits depend strongly on effective EMS coordination under time-varying operating conditions \cite{atef2026integrated,atef2025techno,atef2025real}. Therefore, hydrogen-enabled community microgrids require adaptive EMS frameworks capable of jointly managing renewable intermittency, uncertain demand, short-duration battery operation, and long-duration hydrogen storage.

\subsection{Motivation}
Conventional EMS strategies are commonly based on rule-based control, deterministic scheduling, or offline optimization. Rule-based control is attractive because it is transparent and computationally simple, but fixed dispatch priorities cannot fully respond to rapid changes in renewable output, load demand, electricity prices, or storage availability \cite{Houben2023,Majeed2023,Niknami2024}. Deterministic optimization and metaheuristic methods can improve dispatch quality by solving multi-objective cost-emission or cost-reliability problems, yet their performance depends heavily on forecast accuracy, scenario assumptions, and repeated computation whenever operating conditions change \cite{Tang2024,Asif2024,Bhuvaneshwarri2025}. In renewable-rich community microgrids, these limitations become more pronounced because the EMS must simultaneously coordinate PV, wind, batteries, electrolyzers, hydrogen storage, fuel cells, grid exchange, and uncertain residential demand in near real time.

Recent reviews emphasize that uncertainty-aware EMS methods, forecasting, computational intelligence, and advanced scheduling controllers are becoming essential for microgrids with high renewable penetration and multiple storage technologies \cite{cabrera2022review,sharma2022critical}. Earlier work demonstrated that artificial neural networks can improve the modeling accuracy of hybrid power systems, supporting the broader transition toward data-driven energy-system modeling and control \cite{atef2019utilization}. A subsequent data-driven optimization framework incorporated demand response and generation uncertainty into microgrid energy management, further demonstrating the value of data-informed methods for complex operational decisions \cite{atef2024datadriven}. Broader reviews of computational intelligence and sustainable scheduling controllers also indicate that EMS research is moving toward adaptive controllers that can respond to renewable intermittency, storage dynamics, and demand variability \cite{allwyn2023comprehensive,bilal2024review,mannan2024recent}. A previous Australian real-time EMS study showed that PSO can outperform fixed RBC under time-varying operating conditions, but PSO must be solved repeatedly and does not learn a reusable policy from accumulated operational experience \cite{atef2025real}. This motivates the use of deep reinforcement learning (DRL), where the EMS is formulated as an agent that interacts with the microgrid environment, observes system states, takes dispatch actions, and improves its policy through reward feedback.

Recent DRL-based energy management studies demonstrate why learning-based controllers are suitable for complex sequential decisions. Expert-guided reinforcement learning has been used to improve data efficiency and reliability for electrified-vehicle EMSs, while MPC-assisted DRL has been proposed for fuel-cell/battery systems to improve adaptability, degradation-aware operation, and long-term decision quality \cite{hu2025enhancing,liu2025novel}. In demand-side and building applications, multi-agent DRL and safe DRL frameworks show how learning controllers can manage user comfort, demand response, safety, and energy efficiency under dynamic conditions \cite{abishu2025multi,wang2025safe}. In port microgrids, two-layer DRL has been used to coordinate logistics--energy coupling and battery dispatch under uncertain ship arrivals \cite{song2025two}. In railway and urban rail applications, reward-based and PV-aware DRL methods have addressed regenerative braking, PV fluctuation, and train-load uncertainty \cite{chen2025multi,xu2026drl}. DRL has also been applied to MEC systems with distributed power sources, where task offloading and energy allocation must be coordinated with grid trading \cite{liu2025data}. For hydrogen-enabled community microgrids, DRL is attractive because it can learn adaptive coordination among short-duration battery storage, long-duration hydrogen storage, renewable generation, and grid transactions while accounting for nonlinear device behavior and uncertain operating conditions \cite{upadhyay2024energy,atef2026review}.

\subsection{Research Gap}
Recent microgrid EMS reviews show that the field is moving from conventional rule-based and deterministic optimization methods toward data-driven and uncertainty-aware control, but they also identify gaps in sector-specific deployment and integrated storage coordination \cite{atef2026review,atef2026energy,ma2022review}. Existing DRL applications span vehicles, buildings, ports, railways, and communication networks, whose objectives and constraints differ from those of a residential community microgrid with battery, electrolyzer, hydrogen tank, fuel cell, diesel backup, and grid exchange \cite{hu2025enhancing,liu2025novel,abishu2025multi,wang2025safe,song2025two,chen2025multi,liu2025data,xu2026drl}. The related Australian planning work addresses long-term sizing, while the subsequent real-time study compares RBC and PSO rather than a learned continuous-action policy \cite{atef2026integrated,atef2025techno,atef2025real}. The present contribution is a feasibility-aware application framework that formalizes state-dependent projection of continuous PPO actions for this community-scale battery--hydrogen architecture. It does not alter the PPO learning rule, and its present evaluation remains a single-seed, in-sample proof of concept.

\begin{table}[!htbp]
\centering
\caption{Positioning relative to representative learning-based EMS studies.}
\label{tab:literature_comparison}
\footnotesize
\setlength{\tabcolsep}{3.5pt}
\renewcommand{\arraystretch}{1.08}
\begin{tabular}{
    >{\raggedright\arraybackslash}p{2.00cm}
    >{\raggedright\arraybackslash}p{1.70cm}
    >{\raggedright\arraybackslash}p{3.00cm}
    >{\raggedright\arraybackslash}p{2.25cm}
    >{\raggedright\arraybackslash}p{3.10cm}}
\toprule
Study & Sector & Assets and method & Uncertainty focus & Evaluation scope \\
\midrule
Upadhyay et al. \cite{upadhyay2024energy} & Industrial MG & Distributed sources; optimization-assisted RL & Operational variation & Industrial dispatch case \\
Liu et al. \cite{liu2025novel} & Hybrid power system & Fuel cell and battery; MPC-assisted DRL & Operating-condition variation & Fuel-cell--battery control \\
Wang et al. \cite{wang2025safe} & Building & Building energy systems; safe DRL & Safety and demand variation & Explicit safe-control case \\
Song et al. \cite{song2025two} & Port MG & Port loads and battery; two-layer DRL & Ship-arrival uncertainty & Logistics--energy coupling \\
Xu et al. \cite{xu2026drl} & Urban rail & PV and traction system; DRL & PV and train disturbance & Traction-energy coordination \\
This study & Residential community MG & PV, wind, BESS, H$_2$, DG, grid; projected PPO & Annual chronology and outage probability & 8,760 h in-sample proof of concept \\
\bottomrule
\end{tabular}
\end{table}

\subsection{Contributions}
The main contributions of this paper are summarized as follows:
\begin{itemize}[leftmargin=*]
    \item A feasibility-aware continuous-action PPO EMS is formulated for an Australian community microgrid combining PV, wind, battery storage, hydrogen conversion and storage, diesel backup, and grid exchange.
    \item A mass-consistent hydrogen model and an explicit feasibility mapping are documented for battery, diesel, hydrogen, grid, curtailment, and unmet-load constraints.
    \item Economic, reliability, storage-utilization, and emissions metrics are defined with explicit accounting denominators.
    \item The proposed framework is evaluated using an Australian residential community microgrid case study that extends the previous integrated techno-economic planning framework and real-time EMS studies toward learning-based operation of hybrid PV--wind--battery--hydrogen systems \cite{atef2026integrated,atef2025real}.
    \item The reported experiment is deliberately positioned as an in-sample, single-seed proof of concept; limitations that require new controller benchmarks, unseen trajectories, and duration-based outages are stated explicitly.
\end{itemize}

\subsection{Paper Organization}
The remainder of this paper is organized as follows. Section 2 presents the microgrid architecture, component models, and feasibility constraints. Section 3 formulates the MDP, reward, and PPO controller. Section 4 describes the case study, available data record, parameters, scenarios, and performance metrics. Section 5 presents the training, economic, operational, environmental, and resilience results. Section 6 discusses interpretation, deployment limitations, and required validation, and Section 7 concludes the paper.
\section{System Description}
This section defines the hydrogen-enabled community microgrid considered in this study and formulates the component models required by the proposed energy management system. The system is represented through chronological power-flow relations so that renewable variability, storage dynamics, conversion losses, and grid interactions can be captured at each operating interval.

\subsection{Microgrid Architecture}
The studied microgrid supplies an aggregated residential community and consists of renewable generation, short-term electrical storage, a hydrogen-based long-duration storage pathway, power-conditioning equipment, and an external grid connection. Solar photovoltaic (PV) arrays and wind turbines form the main renewable generation sources. A battery energy storage system (BESS) absorbs short-term renewable surplus and supports fast balancing during load-generation mismatch. The hydrogen subsystem includes an electrolyzer, a hydrogen storage tank, and a fuel cell. When renewable generation exceeds the immediate demand and battery-charging requirements, the electrolyzer can convert part of the surplus electrical energy into hydrogen. During prolonged renewable shortages, the stored hydrogen can be reconverted into electricity through the fuel cell.

The microgrid may also exchange power with the utility grid when the interconnection is available. Grid imports are used to cover residual deficits after local generation and storage dispatch, while grid exports allow unused renewable energy to be sold or transferred outside the community boundary. In this structure, the energy management system coordinates PV and wind generation, battery charging/discharging, electrolyzer operation, fuel-cell dispatch, and grid import/export decisions while respecting device limits and maintaining the power balance.

\begin{figure}[!htbp]
\centering
\begin{tikzpicture}[
    >=Latex,
    scale=0.90,
    transform shape,
    every node/.style={font=\small},
    block/.style={draw=black!75, rounded corners=2pt, line width=0.65pt,
        minimum width=2.35cm, minimum height=0.82cm, align=center, fill=blue!5},
    source/.style={block, fill=green!12},
    storage/.style={block, fill=orange!14},
    load/.style={block, fill=blue!9},
    busblock/.style={block, minimum width=2.9cm, minimum height=1.02cm,
        fill=yellow!18, font=\small\bfseries},
    control/.style={block, fill=purple!12, minimum width=3.6cm,
        font=\small\bfseries},
    electrical/.style={-{Latex[length=2.4mm,width=1.7mm]},
        draw=blue!70!black, line width=0.95pt},
    bielectrical/.style={{Latex[length=2.4mm,width=1.7mm]}-{Latex[length=2.4mm,width=1.7mm]},
        draw=blue!70!black, line width=0.95pt},
    hydrogen/.style={-{Latex[length=2.4mm,width=1.7mm]},
        draw=orange!85!black, line width=1.05pt},
    signal/.style={-{Latex[length=2.1mm,width=1.4mm]},
        densely dashed, draw=purple!75!black, line width=0.75pt}
]
\node[source] (pv) at (-5.0,2.2) {PV array};
\node[source] (wind) at (-5.0,0.95) {Wind turbines};
\node[block] (diesel) at (-5.0,-0.30) {Diesel backup};
\node[block] (grid) at (-5.0,-1.55) {Utility grid};
\node[busblock] (bus) at (0,0.45) {Community AC bus};
\node[load] (load) at (5.0,0.45) {Residential load};
\node[storage] (battery) at (0,2.2) {Battery storage};
\node[storage] (ely) at (-1.25,-2.15) {Electrolyzer};
\node[storage] (tank) at (2.0,-2.15) {H$_2$ tank};
\node[storage] (fc) at (5.0,-2.15) {Fuel cell};
\node[control] (ems) at (0,4.05) {PPO energy-management system};

\coordinate (busw1) at ($(bus.west)+(0,0.30)$);
\coordinate (busw2) at ($(bus.west)+(0,0.10)$);
\coordinate (busw3) at ($(bus.west)+(0,-0.10)$);
\coordinate (busw4) at ($(bus.west)+(0,-0.30)$);
\coordinate (buss1) at ($(bus.south)+(-0.55,0)$);
\coordinate (buss2) at ($(bus.south)+(0.55,0)$);

\draw[electrical] (pv.east) -- (busw1);
\draw[electrical] (wind.east) -- (busw2);
\draw[electrical] (diesel.east) -- (busw3);
\draw[bielectrical] (grid.east) -- (busw4);
\draw[bielectrical] (battery.south) -- (bus.north);
\draw[electrical] (bus.east) -- (load.west);
\draw[electrical] (buss1) -- ++(0,-0.48) -| (ely.north);
\draw[hydrogen] (ely.east) -- (tank.west);
\draw[hydrogen] (tank.east) -- (fc.west);
\draw[electrical] (fc.north) -- ++(0,0.48) -| (buss2);

\draw[signal] (ems.south) -- (battery.north);
\draw[signal] (ems.north west) -- (-6.85,4.47) |- (diesel.west);
\draw[signal] (ems.south west) -- (-7.15,3.63) -- (-7.15,-2.92) -| (ely.south);
\draw[signal] (ems.east) -- (6.90,4.05) -- (6.90,-2.92) -| (fc.south);
\end{tikzpicture}
\caption{
Microgrid architecture and principal electrical, hydrogen, and control flows. Battery and grid power exchange (blue bidirectional arrows), hydrogen flow (orange arrows), and communication and control signals between the PPO-based EMS and the microgrid components (red dashed lines). }
\label{fig:microgrid_architecture}
\end{figure}
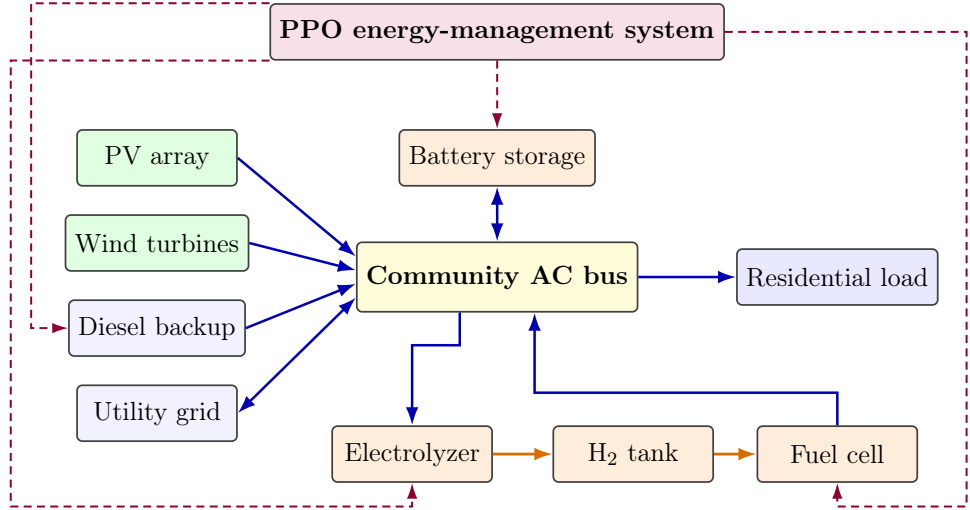

\subsection{Mathematical Models}
The following models describe the main conversion and storage processes over a discrete time interval $\Delta t$. At time step $t$, the electrical load of the community is denoted by $P_{L,t}$. If a representative household demand profile $p_{h,t}$ is available, the community demand is obtained through the aggregation operator
\begin{equation}
P_{L,t}=\sum_{h=1}^{N_h}p_{h,t}=N_{h}p_{h,t},
\label{eq:community_load}
\end{equation}
where $N_{h}$ is the number of households and the second equality applies when all households are represented by the same normalized profile. For a base demand profile $P_{b,t}$ corresponding to $N_{b}$ households, the scaled community load is alternatively expressed through the demand-scaling factor $\alpha_L=N_h/N_b$ as
\begin{equation}
P_{L,t}=\alpha_L P_{b,t},\qquad \alpha_L=\frac{N_h}{N_b}.
\label{eq:scaled_load}
\end{equation}
These relations allow the case-study demand to be represented at the required community scale.

\begin{table}[!htbp]
\centering
\caption{Principal notation and units.}
\label{tab:nomenclature}
\resizebox{\textwidth}{!}{%
\begin{tabular}{lll@{\qquad}lll}
\toprule
Symbol & Meaning & Unit & Symbol & Meaning & Unit \\
\midrule
$P_{L,t}$ & Community demand & kW & $P_{PV,t},P_{WT,t}$ & Available renewable power & kW \\
$P_{B,t}^{ch},P_{B,t}^{dis}$ & Battery charge/discharge power & kW & $SOC_t$ & Battery state of charge & -- \\
$E_B^{rated}$ & Battery energy capacity & kWh & $P_{ely,t}$ & Electrolyzer input power & kW \\
$P_{FC,t}$ & Fuel-cell electrical output & kW & $m_{H,t}$ & Stored hydrogen mass & kg \\
$\dot m_{H,t}^{prod}$ & Hydrogen production flow & kg/h & $\dot m_{H,t}^{cons}$ & Hydrogen consumption flow & kg/h \\
$h_{H_2}^{LHV}$ & Hydrogen lower heating value & kWh/kg & $P_{DG,t}$ & Diesel electrical output & kW \\
$P_{grid,t}^{buy/sell}$ & Grid import/export power & kW & $P_t^{curt}$ & Curtailed renewable power & kW \\
$P_t^{unmet}$ & Unserved load & kW & $\pi_t^{buy/sell}$ & Grid purchase/export tariff & A\$/kWh \\
$G_t$ & Grid availability (1 available) & -- & $H_t$ & Normalized H$_2$ tank level & -- \\
$g_t^{irr}$ & Normalized solar irradiance & -- & $\Delta t$ & Dispatch interval & h \\
\bottomrule
\end{tabular}}
\end{table}

\subsubsection{PV Model}
The PV output is computed from the installed PV capacity, irradiance level, derating effects, and module temperature influence using the standard irradiance- and temperature-corrected PV representation \cite{zia2022energy,Witharama2024}:
\begin{equation}
P_{PV,t}=P_{PV}^{rated} f_{PV}\,g_t^{irr}
\left[1+\gamma_{PV}\left(T_{c,t}-T_c^{ref}\right)\right],
\label{eq:pv_model_rewritten}
\end{equation}
where $P_{PV,t}$ is the PV power output, $P_{PV}^{rated}$ is the installed PV capacity, $f_{PV}$ is the PV derating factor, $g_t^{irr}=I_t/I_{ref}$ is the normalized irradiance, $\gamma_{PV}$ is the temperature coefficient of PV power, $T_{c,t}$ is the cell temperature, and $T_c^{ref}$ is the reference cell temperature. The cell temperature is estimated from ambient temperature and irradiance using
\begin{equation}
T_{c,t}=T_{amb,t}+\left(\frac{T_{NOCT}-T_{amb}^{NOCT}}{I_{NOCT}}\right)I_t,
\label{eq:pv_temperature_rewritten}
\end{equation}
where $T_{amb,t}$ is the ambient temperature, $T_{NOCT}$ is the nominal operating cell temperature, $T_{amb}^{NOCT}=\SI{20}{\celsius}$, and $I_{NOCT}=\SI{800}{\watt\per\square\meter}$.

\subsubsection{Wind Turbine Model}
The wind turbine output is represented using the aerodynamic power relation, with practical dispatch limited by the turbine power curve and rated capacity \cite{zia2022energy}:
\begin{equation}
P_{WT,t}=\frac{\rho C_p A_s}{2}v_t^3,
\label{eq:wind_model_rewritten}
\end{equation}
where $P_{WT,t}$ is the wind power output, $\rho$ is the air density, $A_s$ is the swept rotor area, $C_p$ is the power coefficient, and $v_t$ is the hub-height wind speed. In simulation, this expression is interpreted together with cut-in, rated, and cut-out wind-speed limits to avoid unrealistic operation outside the feasible turbine range.

\subsubsection{Battery SOC Dynamics}
The BESS state of charge evolves according to the charging and discharging power scheduled by the energy management system, following conventional storage energy-balance modeling \cite{yang2022modelling}. A compact single-step expression is
\begin{equation}
SOC_{t+1}=SOC_t+\frac{\Delta t}{E_B^{rated}}
\left(\eta_{ch}P_{B,t}^{ch}-\frac{P_{B,t}^{dis}}{\eta_{dis}}\right),
\label{eq:battery_soc_rewritten}
\end{equation}
where $SOC_t$ is the battery state of charge, $P_{B,t}^{ch}$ and $P_{B,t}^{dis}$ are the battery charging and discharging powers, $\eta_{ch}$ and $\eta_{dis}$ are the charging and discharging efficiencies, and $E_B^{rated}$ is the rated battery energy capacity. The allowable operating range is constrained by
\begin{equation}
\underline{SOC}\leq SOC_t\leq \overline{SOC}.
\label{eq:battery_soc_limits_rewritten}
\end{equation}
This constraint prevents excessive charging or deep discharging and preserves feasible battery operation.

\subsubsection{Hydrogen Subsystem}
The hydrogen subsystem distinguishes tank mass (kg) from hydrogen flow (kg/h). The conversion basis is the hydrogen lower heating value $h_{H_2}^{LHV}=33.33~\mathrm{kWh/kg}$ \cite{modu2023systematic}. The electrolyzer production flow is
\begin{equation}
\dot m_{H,t}^{prod}=\frac{\eta_{ely}P_{ely,t}}{h_{H_2}^{LHV}},
\label{eq:hydrogen_production_rewritten}
\end{equation}
where $\eta_{ely}$ is the LHV-based electrolyzer efficiency. The fuel-cell consumption flow is
\begin{equation}
\dot m_{H,t}^{cons}=\frac{P_{FC,t}}{\eta_{fc}h_{H_2}^{LHV}},
\label{eq:fuel_cell_rewritten}
\end{equation}
where $\eta_{fc}$ is the LHV-based fuel-cell efficiency. The tank state evolves as
\begin{equation}
m_{H,t+1}=m_{H,t}+\left(\dot m_{H,t}^{prod}-\dot m_{H,t}^{cons}\right)\Delta t,
\label{eq:hydrogen_storage_rewritten}
\end{equation}
subject to
\begin{equation}
\underline{m}_H\leq m_{H,t}\leq \overline{m}_H,\quad
0\leq \dot m_{H,t}^{prod}\leq \overline{\dot m}_{H}^{prod},\quad
0\leq \dot m_{H,t}^{cons}\leq \overline{\dot m}_{H}^{cons}.
\label{eq:hydrogen_limits_rewritten}
\end{equation}
The adopted limits are $\overline m_H=1{,}370$~kg, $\overline{\dot m}_H^{prod}=3.90$~kg/h, and $\overline{\dot m}_H^{cons}=7.50$~kg/h. At the stated efficiencies, these flow limits correspond to approximately 200~kW electrolyzer input and 150~kW fuel-cell output. Compression, drying, and balance-of-plant auxiliary electricity are not modeled separately.

\subsubsection{Power Balance Equation}
For every time step, available renewable generation and all dispatchable supply must balance demand, storage charging, hydrogen production, grid export, renewable curtailment, and any residual unserved load. The complete balance is
\begin{align}
P_{PV,t}+P_{WT,t}+P_{FC,t}+P_{DG,t}+P_{B,t}^{dis}
+P_{grid,t}^{buy}+P_t^{unmet}={}&P_{L,t}+P_{B,t}^{ch}+P_{ely,t}\nonumber\\
&+P_{grid,t}^{sell}+P_t^{curt}.
\label{eq:power_balance_rewritten}
\end{align}
Here, $P_{DG,t}$ is diesel output, $P_t^{curt}$ is curtailed renewable power, and $P_t^{unmet}$ is unserved demand after all feasible supply actions have been applied.

\subsubsection{Feasibility Mapping and Operating Constraints}
The raw PPO commands are mapped to a feasible dispatch before the balance in Eq.~\eqref{eq:power_balance_rewritten} is evaluated. Battery power is limited by both the converter rating and the energy available within the SOC bounds:
\begin{align}
0\leq P_{B,t}^{ch}&\leq \min\left\{\overline P_B,\frac{(\overline{SOC}-SOC_t)E_B^{rated}}{\eta_{ch}\Delta t}\right\},\\
0\leq P_{B,t}^{dis}&\leq \min\left\{\overline P_B,\frac{(SOC_t-\underline{SOC})E_B^{rated}\eta_{dis}}{\Delta t}\right\},\\
P_{B,t}^{ch}P_{B,t}^{dis}&=0.
\end{align}
The single signed battery action enforces the final non-simultaneity condition. Hydrogen production and consumption are similarly restricted by tank headroom and inventory:
\begin{align}
0\leq \dot m_{H,t}^{prod}&\leq \min\left\{\overline{\dot m}_H^{prod},\frac{\overline m_H-m_{H,t}}{\Delta t}\right\},\\
0\leq \dot m_{H,t}^{cons}&\leq \min\left\{\overline{\dot m}_H^{cons},\frac{m_{H,t}-\underline m_H}{\Delta t}\right\},\\
\dot m_{H,t}^{prod}\dot m_{H,t}^{cons}&=0.
\end{align}
Let $G_t=1$ when the utility grid is available and $G_t=0$ during an outage. Grid and diesel operation satisfy
\begin{align}
0\leq P_{grid,t}^{buy}&\leq G_t\overline P_{grid}^{buy}, &
0\leq P_{grid,t}^{sell}&\leq G_t\overline P_{grid}^{sell},\\
P_{grid,t}^{buy}P_{grid,t}^{sell}&=0, &
0\leq P_{DG,t}&\leq (1-G_t)\overline P_{DG}.
\end{align}
The reported simulations did not impose finite network import or export ratings; the grid bounds were therefore non-binding whenever $G_t=1$. Residual surplus after feasible charging, hydrogen production, and export is assigned to $P_t^{curt}\geq0$, while residual deficit is assigned to $P_t^{unmet}\geq0$. No cyclic terminal constraints were imposed on $SOC_T$ or $m_{H,T}$, so annual results may depend on the initial and final storage states.

\section{DRL-Based Energy Management Framework}
This section formulates the hourly dispatch problem as a Markov decision process (MDP) and describes the state, action, reward, and PPO learning procedure.
\subsection{DRL Formulation}
The dispatch problem is represented by $\mathcal{M}=(\mathcal{S},\mathcal{A},p,r,\gamma)$. At each hour, the agent observes demand, renewable availability, prices, storage states, time, and grid availability; selects battery, diesel, and hydrogen commands; and receives the next state and reward after the environment balances residual power. The policy $\pi_\theta$ maximizes
\begin{equation}
J(\theta)=\mathbb{E}_{\pi_\theta}\left[\sum_{t=0}^{T-1}\gamma^t r_t\right].
\label{eq:ppo_return}
\end{equation}

\subsection{State Space}
The 11-dimensional observation is
\begin{multline}
s_t=\bigl[SOC_t,H_t,\tau_t,\sin(2\pi\tau_t),\cos(2\pi\tau_t),
\widetilde{P}_{PV,t},\widetilde{P}_{WT,t},\\
\widetilde{P}_{L,t},\widetilde{\pi}^{buy}_t,
\widetilde{\pi}^{sell}_t,G_t\bigr]^{\top}.
\end{multline}
Here $SOC_t$ and $H_t$ are normalized storage levels, $\tau_t$ is normalized hour of day, tildes denote min--max-normalized data, and $G_t$ is the grid-availability indicator.
In the implementation convention used here, $G_t=1$ denotes grid availability and $G_t=0$ denotes an outage. The observation contains current measurements only: it has no day-of-year encoding, forecasts, outage-duration information, or recurrent history. The raw hour variable is also redundant with its sine--cosine encoding. These limitations restrict the evidence for long-horizon or seasonal hydrogen scheduling.
The observation vector is constructed from measurements obtained from the microgrid components, including PV and wind generation, battery state of charge, hydrogen subsystem states (electrolyzer, hydrogen tank, and fuel cell), diesel generator status, load demand, electricity prices, and grid availability. Based on these observations, the PPO-based EMS generates dispatch commands for the controllable resources, thereby establishing the closed-loop interaction between the controller and the microgrid.

\subsection{Action Space}
The continuous action vector is
\begin{equation}
a_t=[a_{B,t},a_{DG,t},a_{H,t}]^{\top},\quad
a_{B,t},a_{H,t}\in[-1,1],\quad a_{DG,t}\in[0,1].
\end{equation}
The feasibility-aware controller applies a state-dependent projection
\begin{equation}
\widehat a_t=\Pi_{\mathcal{F}(s_t)}(a_t),
\end{equation}
where $\mathcal{F}(s_t)$ is the feasible operating set defined by the battery, hydrogen, diesel, grid, and storage-state constraints. The projected command $\widehat a_t$, rather than the raw Gaussian action $a_t$, is passed to the power-balance calculation.
The battery command is scaled by \SI{530}{kW}; positive values discharge and negative values charge. The diesel command is scaled by \SI{180}{kW} and is active only during outages. Positive hydrogen actions operate the electrolyzer and negative actions dispatch the fuel cell. Grid exchange is calculated from residual power balance.
The positive hydrogen branch is scaled to the 3.90~kg/h electrolyzer limit (approximately 200~kW input), whereas the negative branch is scaled to the 7.50~kg/h fuel-cell withdrawal limit (approximately 150~kW electrical output). The feasibility mapping in Section~2 clips each branch to the contemporaneous SOC, tank, device, and grid-availability limits before residual exchange is calculated.

\subsection{Reward Function}
With $\Delta t=1$~h, the hourly net operating cost and reward are written with explicit units as
\begin{align}
C_t={}&\left(\pi^{buy}_tP^{buy}_{grid,t}-\pi^{sell}_tP^{sell}_{grid,t}\right)\Delta t
+c_B\left(P_{B,t}^{ch}+P_{B,t}^{dis}\right)\Delta t\nonumber\\
&+c_{DG}P_{DG,t}\Delta t+\left(c_H^{prod}\dot m^{prod}_{H,t}+c_H^{cons}\dot m^{cons}_{H,t}\right)\Delta t,\\
r_t={}&-C_t-\lambda_U P^{unmet}_t\Delta t
-\lambda_{SOC}|SOC_t-0.5|-\lambda_H|H_t-0.5|,
\end{align}
where $c_B=0.01$~A\$/kWh, $c_{DG}=0.20$~A\$/kWh, $c_H^{prod}=0.40$~A\$/kg, $c_H^{cons}=2.00$~A\$/kg, $\lambda_U=5$~A\$/kWh, $\lambda_{SOC}=0.02$~A\$ per unit SOC deviation per step, and $\lambda_H=0.01$~A\$ per unit normalized hydrogen deviation per step. Export revenue enters with a negative sign.

\begin{table}[!htbp]
\centering
\caption{Reward components and adopted coefficients.}
\label{tab:reward_components}
\resizebox{\textwidth}{!}{%
\begin{tabular}{llll}
\toprule
Component & Coefficient & Unit & Intended purpose \\
\midrule
Grid purchase/export & $\pi_t^{buy},\pi_t^{sell}$ & A\$/kWh & Market cash flow \\
Battery throughput & $c_B=0.01$ & A\$/kWh & Simplified cycling charge \\
Diesel generation & $c_{DG}=0.20$ & A\$/kWh & Fuel/variable operating charge \\
Hydrogen production & $c_H^{prod}=0.40$ & A\$/kg & Production charge \\
Hydrogen withdrawal & $c_H^{cons}=2.00$ & A\$/kg & Withdrawal/usage charge \\
Unserved energy & $\lambda_U=5$ & A\$/kWh & Reliability penalty \\
SOC centering & $\lambda_{SOC}=0.02$ & A\$/step & Discourage boundary operation \\
H$_2$ centering & $\lambda_H=0.01$ & A\$/step & Discourage boundary operation \\
\bottomrule
\end{tabular}
}
\end{table}
The reward coefficients were not subjected to an ablation study. Consequently, the present experiment does not establish that this manually selected reward vector is optimal or that the storage-centering terms improve lifecycle performance.

\subsection{DRL Algorithm}
The EMS uses PPO, an on-policy actor--critic method that limits excessive policy updates \cite{schulman2017proximal}. With probability ratio $\rho_t=\pi_\theta(a_t|s_t)/\pi_{\theta_{old}}(a_t|s_t)$, PPO maximizes
\begin{equation}
L^{clip}=\mathbb{E}_t\left[\min\left(\rho_t\widehat A_t,
\operatorname{clip}(\rho_t,1-\epsilon,1+\epsilon)\widehat A_t\right)\right],
\end{equation}
where $\widehat A_t$ is the generalized advantage estimate and $\epsilon=0.25$. The Gaussian actor and value critic each use two 256-neuron ReLU layers.

\subsection{Training Environment}
Training used 300 episodes of 8,760 hourly steps. Initial storage levels were sampled between 20\% and 80\%. The discount factor was 0.995, GAE factor 0.95, experience horizon 4,096, update epochs 10, mini-batch size 256, entropy weight 0.001, and random seed 42. Reusing the annual chronology makes the baseline evaluation in-sample, while the single seed limits reproducibility. The available evaluation metadata do not identify the checkpoint, checkpoint-selection rule, initial evaluation states, or whether mean or sampled Gaussian actions produced the annual tables.
The proposed DRL-based EMS is implemented using PPO as its core learning and decision-making algorithm. In this framework, the EMS represents the overall control architecture responsible for observing the microgrid operating conditions, determining dispatch actions, enforcing operational constraints through the feasibility mapping, and interacting with the microgrid environment. PPO serves as the optimization algorithm within the EMS, learning an effective dispatch policy through continuous interaction with the environment and reward-based policy updates. Figure 2 illustrates this interaction between the PPO agent and the microgrid environment within the proposed EMS framework.

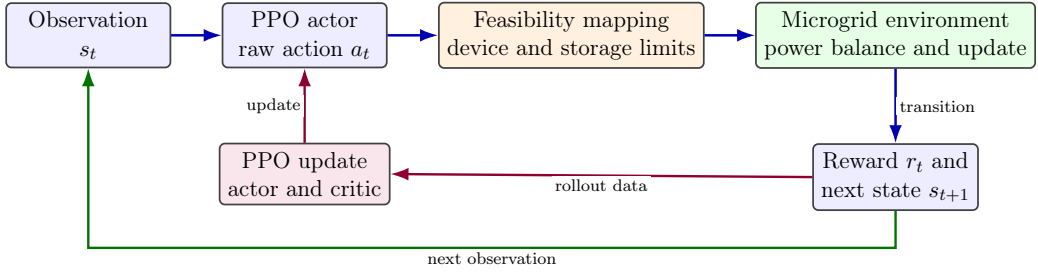
\begin{figure}[!htbp]
\centering
\begin{tikzpicture}[
    >=Latex,
    scale=0.79,
    transform shape,
    node distance=1.25cm and 0.85cm,
    every node/.style={font=\small},
    block/.style={draw=black!75, rounded corners=2pt, line width=0.65pt,
        minimum width=2.75cm, minimum height=1.02cm, align=center, fill=blue!7},
    arrow/.style={-{Latex[length=2.4mm,width=1.7mm]},
        draw=blue!70!black, line width=0.95pt},
    learning/.style={-{Latex[length=2.4mm,width=1.7mm]},
        draw=purple!75!black, line width=0.9pt},
    feedback/.style={-{Latex[length=2.4mm,width=1.7mm]},
        draw=green!45!black, line width=0.9pt},
    flowlabel/.style={font=\scriptsize, fill=white, inner sep=1.5pt}
]
\node[block] (obs) {Observation\\$s_t$};
\node[block, right=of obs] (actor) {PPO actor\\raw action $a_t$};
\node[block, right=of actor, fill=orange!12] (safe) {Feasibility mapping\\device and storage limits};
\node[block, right=of safe, fill=green!10] (env) {Microgrid environment\\power balance and update};
\node[block, below=of env] (reward) {Reward $r_t$ and\\next state $s_{t+1}$};
\node[block, below=of actor, fill=purple!10] (update) {PPO update\\actor and critic};
\draw[arrow] (obs) -- (actor);
\draw[arrow] (actor) -- (safe);
\draw[arrow] (safe) -- (env);
\draw[arrow] (env) -- node[flowlabel, right] {transition} (reward);
\draw[learning] (reward.west) -- node[flowlabel, below] {rollout data} (update.east);
\draw[learning] (update.north) -- node[flowlabel, left] {update} (actor.south);
\draw[feedback] (reward.south) -- ++(0,-0.62) -|
    node[flowlabel, near start, below] {next observation} (obs.south);
\end{tikzpicture}
\caption{PPO interaction and constrained dispatch workflow. During evaluation, the policy update branch is disabled.}
\label{fig:ppo_workflow}
\end{figure}


\section{Case Study and Simulation Setup}
This section describes the Australian residential community microgrid used to evaluate the proposed DRL-based EMS. The case study follows the planning boundary adopted in previous Australian microgrid studies, but the operating problem is reformulated here for learning-based hourly dispatch. The model represents a 1,000-household community rather than the smaller 100-household model used in earlier real-time dispatch testing.

\subsection{Australian Community Case Study}
The case study considers a grid-connected residential community microgrid located in Rockhampton, Queensland, Australia. Rockhampton is selected because it represents a regional Australian community with strong solar potential, available wind resources, and practical interest in resilient distributed energy systems. The community is modeled as an aggregated residential load supplied by local renewable generation, short-duration battery storage, hydrogen storage, dispatchable backup capability, and grid exchange. This boundary is consistent with the previous integrated techno-economic planning framework, which evaluated hybrid PV--wind--battery--diesel--grid--hydrogen configurations for an Australian 1,000-household residential community \cite{atef2026integrated,atef2025techno}.

To avoid ambiguity in the load scale, the present DRL study explicitly adopts the 1,000-household model. If $P_{hh,t}$ denotes the representative per-household demand at time step $t$, the aggregated community demand is defined as
\begin{equation}
P_{L,t}=N_{hh}P_{hh,t}, \qquad N_{hh}=1{,}000,
\label{eq:section4_load_aggregation}
\end{equation}
where $P_{L,t}$ is the total community electrical demand and $N_{hh}$ is the number of households. When a base demand profile is available for a different household count $N_{base}$, the profile is scaled as
\begin{equation}
P_{L,t}=\left(\frac{1{,}000}{N_{base}}\right)P_{base,t},
\label{eq:section4_load_scaling}
\end{equation}
where $P_{base,t}$ is the base aggregated demand. This scaling preserves the chronological shape of residential demand while ensuring that all energy, storage, and grid-exchange quantities correspond to the 1,000-household community scale.

The microgrid architecture includes photovoltaic generation, wind generation, a battery energy storage system, an electrolyzer, hydrogen storage, a fuel cell, grid import/export, and optional backup generation. PV and wind units supply the community directly whenever renewable energy is available. The battery absorbs short-term renewable surplus and supplies rapid demand deficits. The electrolyzer converts excess electricity into hydrogen, the hydrogen tank stores this energy for longer durations, and the fuel cell converts stored hydrogen back to electricity during renewable shortages or high-price periods. The grid interface allows imports when local generation and storage are insufficient and exports when local surplus remains after storage decisions. This configuration provides a suitable testbed for evaluating whether a DRL controller can coordinate fast battery dynamics and slower hydrogen storage dynamics under uncertain renewable, demand, and tariff conditions.

\subsection{Input Data}
The dataset contains 8,760 hourly observations of community demand, available PV and wind power, grid purchase price, and export price. Annual demand is 4.106~GWh, equivalent to approximately 4,106~kWh per household.

\begin{table}[!htbp]
\centering
\caption{Summary of the hourly input dataset.}
\label{tab:input_data_summary}
\resizebox{\textwidth}{!}{%
\begin{tabular}{lrrr}
\toprule
Variable & Mean & Maximum & Annual total \\
\midrule
Community load (kW) & 468.75 & 2,134.13 & 4,106,249.99 kWh \\
PV availability (kW) & 1,620.91 & 7,973.49 & 14,199,183.53 kWh \\
Wind availability (kW) & 18.27 & 60.00 & 160,045.69 kWh \\
Grid purchase price (A\$/kWh) & 0.3060 & 0.4571 & -- \\
Grid export price (A\$/kWh) & 0.0601 & 0.0601 & -- \\
\bottomrule
\end{tabular}}
\end{table}

At each time step, the environment updates the available PV power, wind power, load demand, tariff values, battery state of charge, and hydrogen storage level. These variables form the core information stream used by the DRL agent. The use of chronological data also allows storage continuity to be represented correctly, because the battery state of charge and hydrogen storage level at one time step depend on previous charging, discharging, hydrogen-production, and fuel-cell decisions. This is particularly important for the proposed hybrid storage system because batteries and hydrogen tanks operate over different time scales.

Exogenous variables are min--max normalized for the policy input according to
\begin{equation}
\widetilde{x}_t=\frac{x_t-x_{min}}{x_{max}-x_{min}},
\end{equation}
while raw values are retained for power balancing and cost calculation. The available dataset archive contains the hourly aggregate series and summary statistics but not the source year, load provider, weather station, tariff source, cleaning procedure, or missing-data treatment. Software version, computing hardware, training time, inference time, and code/data availability are also unavailable; consequently, independent reproduction is presently limited.

\subsection{Model Parameters and Operating Constraints}
\begin{table}[!htbp]
\centering
\caption{Principal component parameters.}
\label{tab:component_parameters}
\resizebox{\textwidth}{!}{%
\begin{tabular}{lll}
\toprule
Subsystem & Parameter & Value \\
\midrule
Battery & Capacity / maximum rate & 5,300 kWh / 530 kW \\
Battery & Charge / discharge / round-trip efficiency & 0.95 / 0.95 / 0.9025 \\
Diesel & Maximum output & 180 kW \\
Hydrogen tank & Capacity & 1,370 kg \\
Hydrogen conversion & LHV basis & 33.33 kWh/kg \\
Electrolyzer & Maximum production / efficiency / input & 3.90 kg/h / 0.65 / 200 kW \\
Fuel cell & Maximum withdrawal / efficiency / output & 7.50 kg/h / 0.60 / 150 kW \\
Grid & Baseline / stress outage probability & 1\% / 5\% per hour \\
Grid & Import / export rating & No finite limit imposed \\
Terminal state & SOC / hydrogen constraint & None imposed \\
\bottomrule
\end{tabular}}
\end{table}
The PV and wind columns in the hourly dataset are treated as exogenous available-power series. Their maximum values are 7.973~MW and 60~kW, respectively. Installed nameplate capacities and the conversion procedure used to generate these series are not part of the available dataset archive, limiting assessment of sizing and connection eligibility.

\subsection{Simulation Workflow}
At each hour, the agent selects battery, diesel, and hydrogen actions. The environment applies the feasibility mapping, determines residual grid exchange, curtailment, or unmet load, updates storage, and returns the operating reward. The annual comparison contains only the baseline reliability case with 1\% independent hourly outage probability and the 5\% outage-probability stress case. Unmatched high-variability, hydrogen-cost, and random-policy runs are excluded.
The proposed DRL-based EMS, including the PPO controller and the microgrid simulation environment, was implemented in MATLAB. The simulation was performed using an hourly time-step over the 8,760-hour annual dataset, with MATLAB used for controller implementation, environment modeling, and performance evaluation.

\subsection{Hyperparameters}
\begin{table}[!htbp]
\centering
\caption{PPO training hyperparameters.}
\resizebox{\textwidth}{!}{%
\begin{tabular}{lr@{\qquad}lr}
\toprule
Parameter & Value & Parameter & Value \\
\midrule
Episodes & 300 & Steps/episode & 8,760 \\
Discount / GAE factor & 0.995 / 0.95 & Experience horizon & 4,096 \\
Epochs / mini-batch & 10 / 256 & Clip / entropy weight & 0.25 / 0.001 \\
Actor / critic & $2\times256$ & State / action size & 11 / 3 \\
Random seed & 42 & Training chronologies & One annual profile \\
\bottomrule
\end{tabular}}
\end{table}
The reported PPO configuration is incomplete because actor and critic learning rates, optimizer settings, gradient clipping, value-loss coefficient, policy-standard-deviation bounds, and checkpointing rules were not preserved with the aggregate results.

\subsection{Evaluation Scenarios}
The baseline reliability case uses a 1\% independent hourly grid-outage probability; it is not a no-outage case and is therefore not described as normal operation. The outage stress case increases that probability to 5\%. Because independent Bernoulli sampling primarily produces isolated outage hours, neither case represents sustained outage duration. The hydrogen-cost accounting sensitivity raises production cost from A\$0.40 to A\$1.00/kg and withdrawal cost from A\$2.00 to A\$10.00/kg. Reported economics exclude capital, replacement, fixed O\&M, connection charges, and lifecycle expenditure.

\subsection{Performance Metrics}
For any power series $P_{x,t}$, annual energy is $E_x=\sum_tP_{x,t}\Delta t$. The annual net operating cash balance is the negative of accumulated operating cost,
\begin{equation}
B_{op}=-\sum_t C_t.
\end{equation}
Load satisfaction and net grid exchange are
\begin{align}
LS&=100\left(1-\frac{E_{unmet}}{E_L}\right),\\
E_{grid}^{net}&=E_{grid}^{buy}-E_{grid}^{sell}.
\end{align}
The existing renewable indicator is renamed the gross renewable share of primary electrical supply (GRS):
\begin{equation}
GRS=100\frac{E_{PV}+E_{WT}}{E_{PV}+E_{WT}+E_{grid}^{buy}+E_{DG}}.
\end{equation}
GRS is not the fraction of community demand supplied by renewables because it includes renewable energy exported outside the community. A community renewable self-sufficiency value would require source tracing through battery and hydrogen charging and is not recoverable from the aggregate annual outputs.

Let $M_{tot}$ denote grid-import, diesel, PV-lifecycle, and wind-lifecycle emissions. The separate hydrogen-production term used in the earlier calculation is excluded because it applied another electricity-related factor to electrolyzer input after the source electricity had already been counted. Define served community energy, export-inclusive delivered energy, and primary electrical supply as
\begin{align}
E_{served}&=E_L-E_{unmet},\\
E_{del}&=E_{served}+E_{grid}^{sell},\\
E_{primary}&=E_{PV}+E_{WT}+E_{grid}^{buy}+E_{DG}.
\end{align}
The three reported carbon-intensity indicators are
\begin{equation}
CI_{served}=\frac{M_{tot}}{E_{served}},\qquad
CI_{del}=\frac{M_{tot}}{E_{del}},\qquad
CI_{gen}=\frac{M_{tot}}{E_{primary}}.
\end{equation}
$CI_{served}$ conservatively assigns the full system inventory to served community demand, $CI_{del}$ includes community delivery and exports, and $CI_{gen}$ normalizes by external and renewable electricity entering the system before storage conversion. None of these denominators counts battery or fuel-cell output as a new primary energy input. The earlier avoided-emissions indicator is removed because its one-for-one export-displacement assumption and internally double-counted denominator were not sufficiently defensible. The economic resilience factor and battery discharge-to-charge ratio are
\begin{equation}
R_{econ}=\frac{B_{op}^{stress}}{B_{op}^{base}},\qquad
R_B=\frac{E_B^{dis}}{E_B^{ch}}.
\end{equation}
Because no cyclic terminal condition is imposed, $R_B$ is a utilization indicator rather than an equivalent-cycle metric.
\section{Results}
This section presents the single-seed PPO training record and the economic, operational, environmental, and resilience results for the 8,760-hour community chronology. Negative operating cost denotes a positive net operating cash balance. The main comparison is limited to the baseline reliability and 5\% outage-probability cases; unmatched diagnostic runs are excluded.

\subsection{DRL Training Performance}
The agent completed 300 annual episodes, corresponding to 2.628 million environment steps. The best recorded training reward was $-8{,}937.58$ at Episode~95. A sharp policy collapse occurred at Episode~97, when the reward decreased to $-1{,}246{,}934.13$, followed by a minimum of $-1{,}400{,}100.39$ at Episode~115. Recovery became pronounced after Episode~210, and the final reward reached $-22{,}078.98$. The original per-episode numerical series and moving-average specification are unavailable, so the earlier raster training plot is not reproduced. Because no validation chronology or multi-seed record is available and the evaluated checkpoint is unidentified, Episode~95 is reported only as the best training return, not as a selected best model.

\begin{table}[!htbp]
\centering
\caption{PPO training summary.}
\begin{tabular}{lr}
\toprule
Metric & Value \\
\midrule
Episodes / steps & 300 / 2,628,000 \\
Best reward & $-8{,}937.58$ (Episode 95) \\
Final reward & $-22{,}078.98$ \\
Final-window average & $-22{,}156.79$ \\
Final initial-state value, $Q_0$ & $-528.27$ \\
\bottomrule
\end{tabular}
\end{table}

\begin{table}[!htbp]
\centering
\caption{Observed PPO degradation and recovery pattern.}
\begin{tabular}{lll}
\toprule
Episode range & Reward development & Interpretation \\
\midrule
97--115 & $-1.25$ to $-1.40$ million & Policy degradation \\
116--150 & $-1.38$ to $-1.31$ million & Limited recovery \\
151--210 & $-1.38$ to $-0.78$ million & Progressive recovery \\
211--300 & $-0.65$ to $-0.022$ million & Rapid restabilization \\
\bottomrule
\end{tabular}
\end{table}

The reward standard deviation decreased from 294,847.28 over Episodes~1--96 to 3,847.62 over Episodes~211--300. The lower late-stage dispersion indicates restabilization after recovery, but the single trajectory does not establish repeatable convergence. The large collapse and the gap between the best and final rewards motivate lower clipping values, explicit learning-rate and gradient-clipping control, validation-based checkpointing, and at least five independent training seeds.

\paragraph{Critical interpretation.}
The numerical pattern does not demonstrate monotonic PPO convergence. Although late-stage reward variability is much lower, the magnitude of the final negative reward is approximately 2.47 times that of the best training reward. Reduced variance can therefore indicate a stable but inferior policy rather than convergence to the best observed policy. Moreover, the two-order-of-magnitude deterioration after Episode~95 is too large to treat as ordinary exploration noise; plausible causes include an excessive policy update, critic instability, or an interaction between policy updates and the feasibility-projection layer, but the archived outputs are insufficient to identify the cause. Because the evaluated checkpoint is unknown, the annual results cannot be attributed confidently to either the Episode~95 policy or the final policy. The defensible conclusion is that one run recovered from a severe collapse, not that training was reproducibly stable or converged.

\subsection{Economic Performance}
The baseline reliability case produced an annual net operating cash balance of A\$195,690.67, primarily because 11.415~GWh was exported while 1.387~GWh was imported. The 5\% outage-probability case retained A\$169,892.21, corresponding to an economic resilience factor of 0.868 and an annual reduction of A\$25,798.46. These values describe operation of an already installed system; they are not profit, net present value, or evidence of project-level economic viability.

\begin{table}[!htbp]
\centering
\caption{Annual operating-cost breakdown (A\$).}
\resizebox{\textwidth}{!}{%
\begin{tabular}{lrr}
\toprule
Scenario & Baseline reliability & 5\% outage probability \\
\midrule
Net grid transaction & $-205{,}062.01$ & $-186{,}563.48$ \\
Battery throughput charge & 398.26 & 2,157.88 \\
Diesel variable charge & 1,155.53 & 6,117.38 \\
Hydrogen charge & 7,817.55 & 8,396.01 \\
Total operating cost & $-195{,}690.67$ & $-169{,}892.21$ \\
\bottomrule
\end{tabular}}
\end{table}

The net-grid entries include import expenditure and export revenue with their physical signs; this is why they differ from the larger gross grid-transaction totals obtained by summing absolute hourly transactions. The separately seeded hydrogen-cost run is excluded because it is neither a controlled accounting sensitivity nor a price-responsive control experiment.

\paragraph{Critical interpretation.}
The economic outcome is dominated by the grid boundary rather than by storage savings. Baseline exports are approximately 2.78 times annual community demand, and the A\$205,062.01 net grid revenue exceeds the final A\$195,690.67 cash balance because it also pays all modeled battery, diesel, and hydrogen charges. When outage probability increases, exports fall by 471.46~MWh and the net grid contribution deteriorates by A\$18,498.53, accounting for approximately 71.7\% of the total A\$25,798.46 reduction in cash balance. The remaining reduction is mainly associated with higher battery throughput and diesel use. Thus, the apparent economic resilience reflects continued export capability as much as resilient load supply. Because export capacity is non-binding and capital, fixed O\&M, replacement, degradation, and connection costs are omitted, the result is highly sensitive to the assumed export tariff and renewable-system scale and cannot establish lifecycle profitability or the economic superiority of PPO.

\subsection{Operational Performance}
The baseline reliability case exported 11.415~GWh and curtailed 103.76~MWh, equivalent to approximately 0.72\% of available PV and wind generation. With 5\% independent hourly outage probability, battery discharge increased by 412.9\%, battery charging by 471.6\%, diesel generation by 429.4\%, and hydrogen electrical output by 7.7\%. The larger battery and diesel responses show that these resources carried most of the isolated-hour outage burden, whereas hydrogen supplied a smaller incremental contribution.

\begin{figure}[!htbp]
\centering
\includegraphics[width=0.94\textwidth]{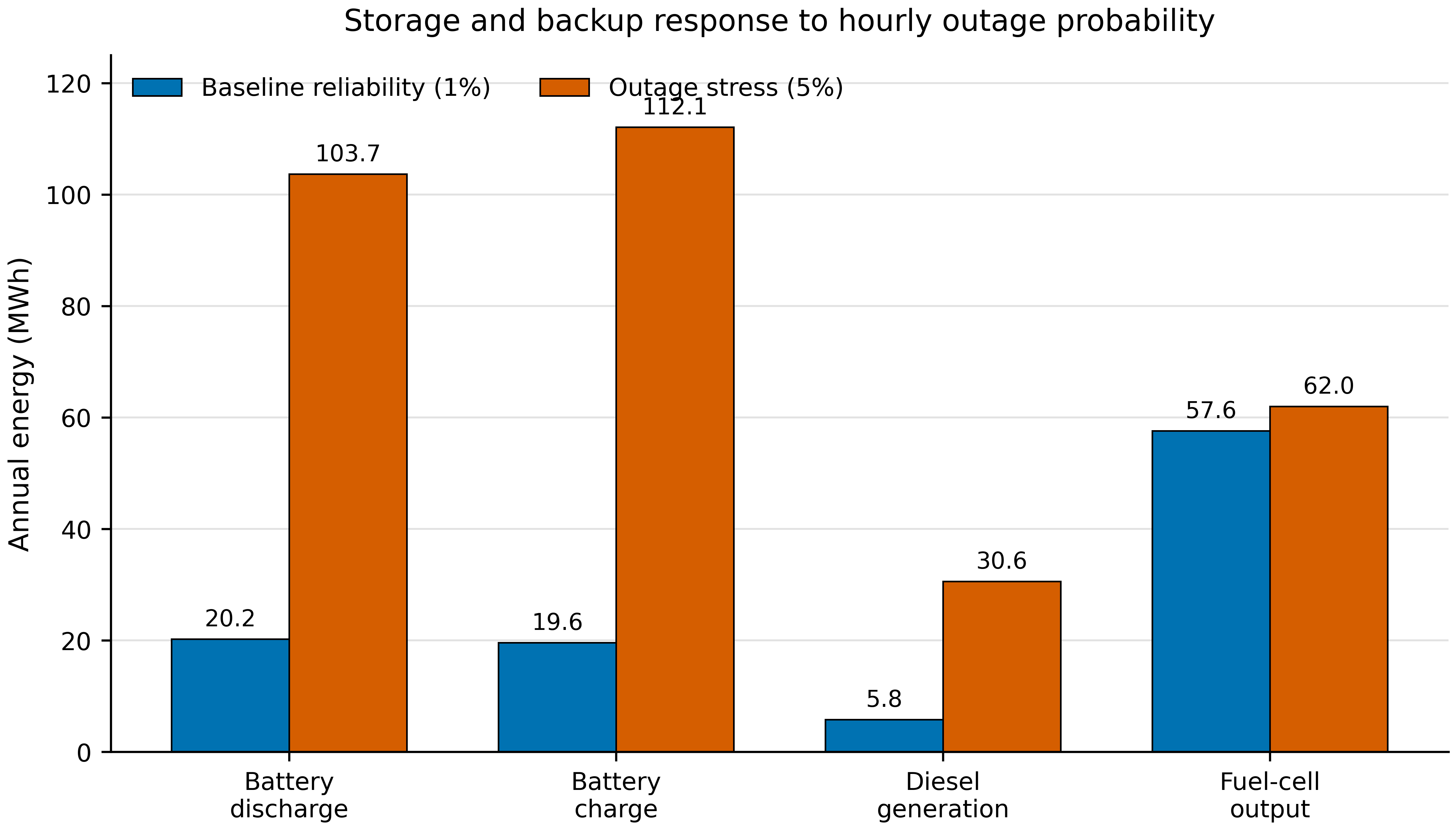}
\caption{Storage and backup response when hourly outage probability increases from 1\% to 5\%.}
\end{figure}

\begin{table}[!htbp]
\centering
\caption{Operational energy balance (kWh).}
\resizebox{\textwidth}{!}{%
\begin{tabular}{lrr}
\toprule
Quantity & Baseline reliability & 5\% outage probability \\
\midrule
PV / wind & 14,199,183.53 / 160,045.69 & 14,199,183.53 / 160,045.69 \\
Battery discharge / charge & 20,215.52 / 19,610.00 & 103,687.77 / 112,100.03 \\
Diesel generation & 5,777.66 & 30,586.90 \\
Grid import / export & 1,386,690.91 / 11,414,873.52 & 1,359,124.73 / 10,943,413.24 \\
H$_2$ output / production input & 57,574.17 / 182,013.41 & 61,981.06 / 193,576.18 \\
Unmet load & 9,254.04 & 49,585.27 \\
Renewable curtailed & 103,759.99 & 578,002.91 \\
\bottomrule
\end{tabular}}
\end{table}

Curtailment increased to 578.00~MWh in the outage case, 4.03\% of annual PV and wind availability and 457\% above the baseline reliability case. During grid-unavailable hours, surplus renewable energy cannot be exported, and finite charging and electrolyzer limits prevent complete absorption of that surplus. The battery discharge-to-charge ratio changed from 1.03 to 0.925. Because this ratio is affected by initial and final state of charge, it is interpreted as a utilization indicator rather than an exact equivalent-cycle count.

\paragraph{Critical interpretation.}
The operational totals show a clear hierarchy in flexibility provision. Battery discharge rises by 83.47~MWh and diesel generation by 24.81~MWh, whereas fuel-cell output rises by only 4.41~MWh; the studied outage process therefore activates fast electrical storage and backup generation much more strongly than hydrogen storage. Relative to the 5.3~MWh battery capacity, annual discharge corresponds to approximately 3.8 nominal capacity-throughput equivalents in the baseline case and 19.6 under outage stress, indicating a substantial increase in battery utilization and likely degradation exposure that is not represented by the simplified throughput charge. Curtailment becomes 5.57 times the baseline value despite increased battery charging and electrolyzer input, revealing a temporal bottleneck when grid export is unavailable and conversion limits bind. The annual fuel-cell-output-to-electrolyzer-input ratios of 31.6\% and 32.0\% should not be interpreted as realized round-trip efficiencies because initial and final tank inventories are not reported. More generally, annual aggregates cannot show whether charging and discharging decisions were well timed; hourly trajectories and terminal storage states are required to establish physically and strategically effective coordination.

\subsection{Environmental Performance}
After removing the duplicate hydrogen-electricity term, total modeled emissions are 1.342~kt~CO$_2$ in the baseline reliability case and 1.349~kt~CO$_2$ at 5\% hourly outage probability. Baseline intensities are 0.328~kg~CO$_2$/kWh for served community demand, 0.087~kg~CO$_2$/kWh for export-inclusive delivered energy, and 0.085~kg~CO$_2$/kWh for primary electrical supply. The corresponding outage-stress values are 0.333, 0.090, and 0.086~kg~CO$_2$/kWh.

\begin{table}[!htbp]
\centering
\caption{Emission factors reconstructed from the reported component calculations.}
\label{tab:emission_factors}
\resizebox{\textwidth}{!}{%
\begin{tabular}{llllll}
\toprule
Source & Factor & Unit & Application basis & Boundary treatment & Source status \\
\midrule
Grid electricity & 0.45 & kg CO$_2$/kWh & Imported electricity & Counted once at import & Reconstructed assumption \\
Diesel & 0.80 & kg CO$_2$/kWh & Diesel electrical output & Counted once at generation & Reconstructed assumption \\
PV lifecycle & 0.05 & kg CO$_2$/kWh & Available PV generation & Includes curtailed availability & Reconstructed assumption \\
Wind lifecycle & 0.02 & kg CO$_2$/kWh & Available wind generation & Includes curtailed availability & Reconstructed assumption \\
Hydrogen conversion & -- & -- & Internal electricity conversion & No additional electricity factor & Corrected boundary \\
\bottomrule
\end{tabular}}
\end{table}
These factors reproduce the component totals in the available results, but their external source citations are not contained in the study archive. Separate electrolyzer, tank, and fuel-cell manufacturing emissions are outside the adopted boundary.

\begin{table}[!htbp]
\centering
\caption{Environmental performance and emission-source breakdown.}
\resizebox{\textwidth}{!}{%
\begin{tabular}{lrr}
\toprule
Metric & Baseline reliability & 5\% outage probability \\
\midrule
Grid emissions (kg CO$_2$) & 624,010.91 & 611,606.13 \\
Diesel emissions (kg CO$_2$) & 4,622.13 & 24,469.52 \\
PV lifecycle emissions (kg CO$_2$) & 709,959.18 & 709,959.18 \\
Wind lifecycle emissions (kg CO$_2$) & 3,200.91 & 3,200.91 \\
Total emissions (kg CO$_2$) & 1,341,793.13 & 1,349,235.74 \\
Served-demand intensity, $CI_{served}$ (kg CO$_2$/kWh) & 0.328 & 0.333 \\
Export-inclusive delivered intensity, $CI_{del}$ (kg CO$_2$/kWh) & 0.087 & 0.090 \\
Generation-based intensity, $CI_{gen}$ (kg CO$_2$/kWh) & 0.085 & 0.086 \\
Gross renewable share, GRS (\%) & 91.2 & 91.2 \\
\bottomrule
\end{tabular}}
\end{table}

\begin{figure}[!htbp]
\centering
\includegraphics[width=0.90\textwidth]{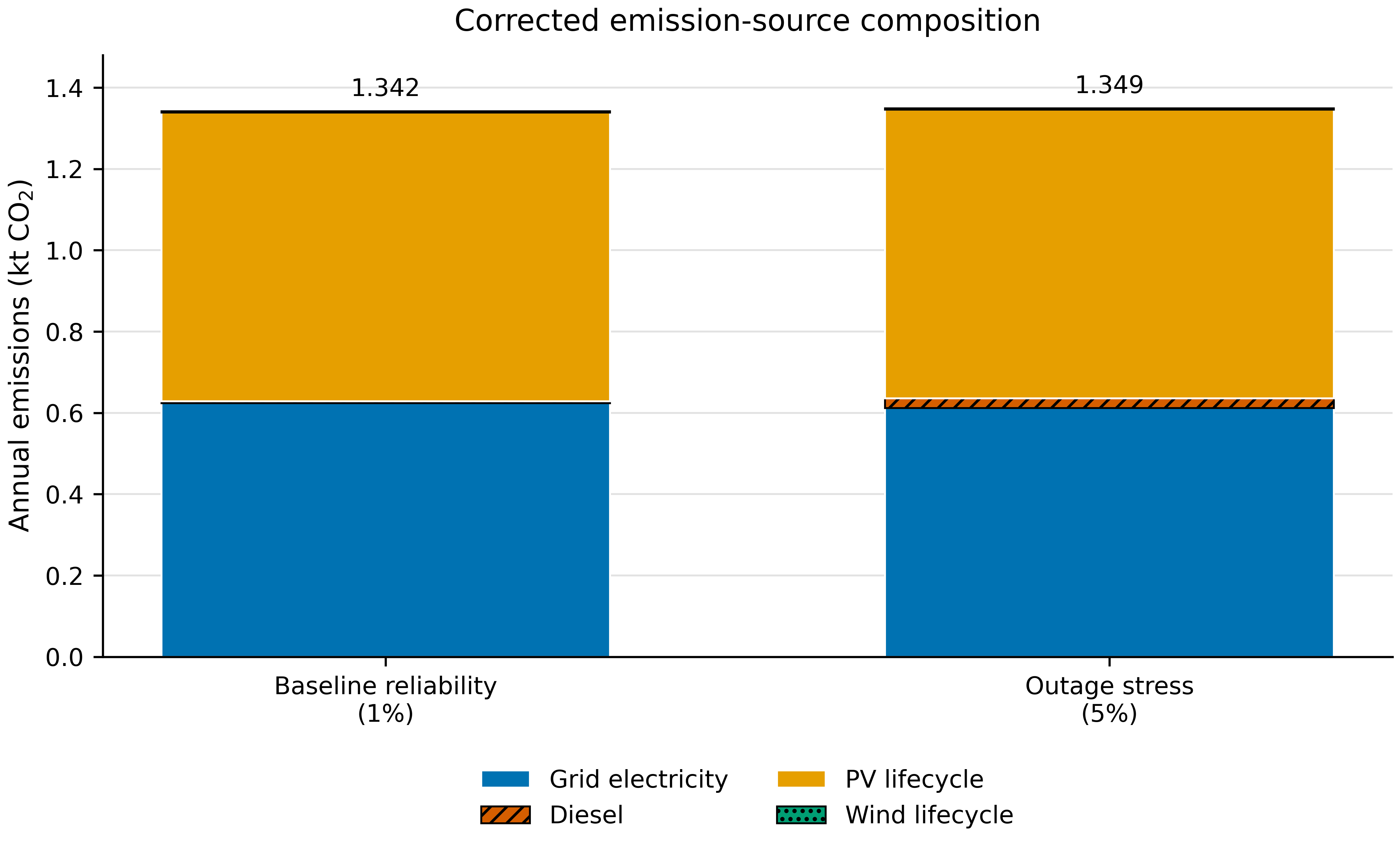}
\caption{Contribution of modeled emission sources to annual emissions.}
\label{fig:emissions_breakdown_full}
\end{figure}

PV lifecycle emissions contribute 52.9\% of the corrected baseline inventory, followed by imported grid electricity at 46.5\%; diesel and wind collectively contribute approximately 0.6\%. No avoided-emissions value is reported because allocating exported renewable electricity to displaced grid generation requires an explicit marginal-emissions and export-allocation model.

\paragraph{Critical interpretation.}
The 5\% outage case increases the corrected inventory by only 7.44~t~CO$_2$, or 0.55\%, but this small net change masks an important substitution: grid-import emissions decrease by 12.40~t while diesel emissions increase by 19.85~t. The environmental penalty is therefore driven by replacing some grid-supported operation with carbon-intensive local backup. The three reported intensities answer different questions. The low export-inclusive value is strongly denominator-driven by more than 11~GWh of exported electricity and must not be interpreted as the carbon intensity of electricity consumed by the community. Conversely, the served-demand intensity conservatively assigns the full PV and wind lifecycle inventory, including generation that is exported or curtailed, to local served demand. Because PV lifecycle emissions are applied to all available PV generation, they are unchanged between scenarios and are controlled mainly by system sizing rather than PPO dispatch. The environmental results are consequently best interpreted as accounting-boundary sensitivities; source-verified emission factors, asset manufacturing boundaries, and an explicit export-allocation method are needed for comparative carbon claims.

\subsection{Resilience Analysis}
At 5\% hourly outage probability, unmet load increased from 9.25 to 49.59~MWh and load satisfaction decreased from 99.77\% to 98.79\%, where both percentages are calculated against the measured annual community demand of 4.106~GWh. The system retained 86.8\% of the baseline net operating cash balance. Earlier percentages based on a different denominator are not used.

\begin{figure}[!htbp]
\centering
\includegraphics[width=\textwidth]{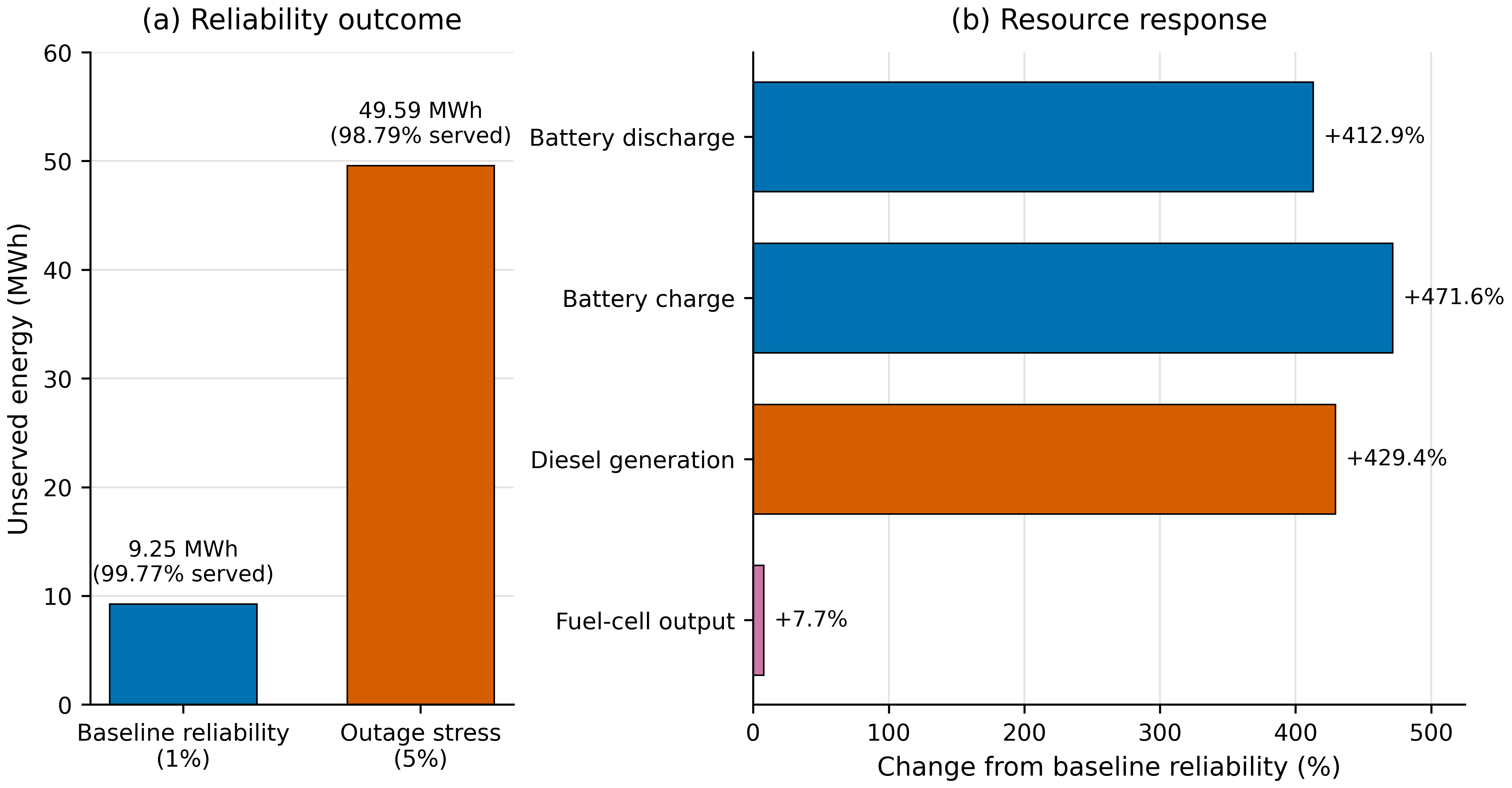}
\caption{Reliability and resource response at 1\% and 5\% independent hourly outage probability.}
\end{figure}

\begin{table}[!htbp]
\centering
\caption{Resilience metrics.}
\resizebox{\textwidth}{!}{%
\begin{tabular}{lrrr}
\toprule
Metric & Baseline reliability & 5\% outage probability & Change \\
\midrule
Net operating cash balance (A\$) & 195,690.67 & 169,892.21 & $-13.2\%$ \\
Unmet load (kWh) & 9,254.04 & 49,585.27 & $+435.8\%$ \\
Load satisfaction (\%) & 99.77 & 98.79 & $-0.98$ points \\
Battery discharge (kWh) & 20,215.52 & 103,687.77 & $+412.9\%$ \\
Battery charge (kWh) & 19,610.00 & 112,100.03 & $+471.6\%$ \\
Diesel generation (kWh) & 5,777.66 & 30,586.90 & $+429.4\%$ \\
H$_2$ production input (kWh) & 182,013.41 & 193,576.18 & $+6.4\%$ \\
H$_2$ electrical output (kWh) & 57,574.17 & 61,981.06 & $+7.7\%$ \\
Hydrogen consumed (kg) & 3,143.55 & 3,384.17 & $+7.7\%$ \\
\bottomrule
\end{tabular}}
\end{table}

The 24.81~MWh increase in diesel generation provided the largest dispatchable backup increment. Battery discharge increased by 83.47~MWh, while hydrogen electrical output increased by only 4.41~MWh. Consequently, the present outage experiment demonstrates effective short-event support from the battery and diesel generator but provides limited evidence for long-duration resilience from hydrogen. Multi-hour and multi-day outage sequences are required to quantify that role.

\paragraph{Critical interpretation.}
The percentage-point change in load satisfaction appears modest, but it corresponds to an additional 40.33~MWh of unserved energy and a 5.36-fold increase in annual unmet load. Reporting only the 98.79\% satisfaction value would therefore understate the practical reliability deterioration. Economic and service resilience also diverge: the system retains 86.8\% of its cash balance while unmet load increases by 435.8\%, because export revenue contributes strongly to the economic metric even when some community demand is not supplied. In addition, the baseline already contains a 1\% independent hourly outage probability, so this experiment compares two reliability-stress levels rather than outage operation against a fully available grid. Independent Bernoulli outages emphasize isolated short events and do not test energy autonomy across sustained failures; this structure also explains why battery and diesel response dominates hydrogen. Realized outage hours, common outage traces, event-duration statistics, critical-load performance, and recovery trajectories are required before assigning a general resilience value to the controller or to hydrogen storage.

\subsection{Comparative Analysis}
Only the baseline reliability and 5\% outage-probability cases share the same annual load and renewable profiles and are retained in the consolidated comparison. No formal controller ranking is presented because an annual RBC, PSO, optimization, SAC, or TD3 benchmark under identical conditions is not available. Likewise, the synthetic high-variability, separately seeded hydrogen-cost, and 1,000-hour random-policy runs are not used as comparative evidence.

\begin{table}[!htbp]
\centering
\caption{Consolidated scenario performance.}
\resizebox{\textwidth}{!}{%
\begin{tabular}{lrr}
\toprule
Metric & Baseline reliability & 5\% outage probability \\
\midrule
Total operating cost (A\$) & $-195{,}690.67$ & $-169{,}892.21$ \\
Unmet load (kWh) & 9,254.04 & 49,585.27 \\
Diesel generation (kWh) & 5,777.66 & 30,586.90 \\
Hydrogen consumed (kg) & 3,143.55 & 3,384.17 \\
Net grid exchange (kWh) & $-10{,}028{,}182.61$ & $-9{,}584{,}288.51$ \\
Gross renewable share, GRS (\%) & 91.2 & 91.2 \\
Served-demand intensity (kg CO$_2$/kWh) & 0.328 & 0.333 \\
Export-inclusive delivered intensity (kg CO$_2$/kWh) & 0.087 & 0.090 \\
\bottomrule
\end{tabular}}
\end{table}

\paragraph{Critical interpretation.}
This table is a controlled scenario sensitivity only, not a controller comparison. It shows that increasing hourly outage probability reduces cash balance, increases unmet load and diesel use, and worsens both demand-normalized and export-inclusive emission intensities. However, it cannot determine whether PPO manages those changes better than an RBC, PSO, deterministic optimizer, SAC, or TD3 controller because no alternative controller was evaluated with the same chronology, outage trace, initial storage states, and accounting rules. The unchanged 91.2\% gross renewable share, despite a 5.36-fold increase in unmet load and a 5.57-fold increase in curtailment, also shows that this annual aggregate is not sufficiently sensitive to characterize operational resilience or renewable utilization. A credible comparative claim requires common-condition benchmarks, multiple PPO seeds, validation-selected checkpoints, finite grid limits, and matched terminal storage conditions.

\section{Discussion}
\subsection{Economic Interpretation}
The positive operating cash balance is dominated by renewable export. Annual exports of 11.415~GWh are approximately 2.8 times the 4.106~GWh community demand, while available PV and wind energy is approximately 3.5 times demand. The result therefore reflects a large renewable availability profile, an export tariff, and a non-binding grid-export limit, not only storage coordination. Capital expenditure, fixed O\&M, replacement, network augmentation, connection charges, and most degradation costs are excluded. Accordingly, A\$195,690.67 is an operating cash balance for an assumed installed system rather than profit or project viability. Installed-capacity documentation and export-constrained cases are necessary for a network-realistic economic assessment.

\subsection{Battery, Hydrogen, and Diesel Roles}
When hourly outage probability increased from 1\% to 5\%, battery discharge increased by 83.47~MWh and diesel generation by 24.81~MWh, whereas fuel-cell electrical output increased by only 4.41~MWh. This pattern is consistent with isolated short events, for which fast battery response and diesel backup are more immediately useful than long-duration conversion through hydrogen. The revised hydrogen equations clarify the mass and LHV conversion boundary, but the experiment still does not demonstrate seasonal or long-duration resilience. Duration-based outages of 6--72~h, low-renewable multi-day events, minimum storage trajectories, and post-event recovery are required to quantify the specific contribution of hydrogen.

\subsection{DRL Performance, Stability, and Generalization}
The Episode~97 collapse, the single training seed, and the unidentified evaluated checkpoint prevent a reproducibility claim. Reusing the training chronology for evaluation also makes the annual results in-sample. Generalization and comparative performance therefore remain untested against unseen trajectories and the RBC and PSO approaches used in related Australian work \cite{atef2025real}. A complete assessment would use validation-based checkpointing, multiple seeds, unseen annual tests, and common-condition controller benchmarks. Seasonal encoding, forecasts, and temporal memory may also improve strategic hydrogen dispatch.

\subsection{Practical Deployment and Remaining Limitations}
Practical deployment would require verified action projection, finite network limits, terminal storage control, degradation and startup models, hydrogen auxiliary consumption, and testing under forecast, communication, and measurement errors. Reproducible data and software metadata and inference-time measurements are also absent. Because related DRL studies address different sectors and accounting boundaries (Table~\ref{tab:literature_comparison}), their absolute results do not establish comparative superiority. Hourly dispatch traces would provide additional evidence of physically meaningful battery and hydrogen behavior.

\section{Conclusion}
This paper presented a feasibility-aware PPO proof of concept for hourly operation of a hydrogen-enabled microgrid serving 1,000 households. In the 1\% outage-probability baseline, the annual net operating cash balance was A\$195,690.67, load satisfaction was 99.77\%, and the gross renewable share was 91.2\%. Corrected modeled emissions were 1.342~kt~CO$_2$, corresponding to 0.328~kg~CO$_2$/kWh of served community demand and 0.087~kg~CO$_2$/kWh of export-inclusive delivered energy. At 5\% hourly outage probability, the cash balance decreased to A\$169,892.21 and load satisfaction to 98.79\%, with battery and diesel use increasing much more than fuel-cell output. The study demonstrates constraint-feasible dispatch on the examined chronology; broader conclusions require unseen testing, common-condition benchmarks, multiple seeds, finite export limits, terminal storage control, and sustained-outage cases.

\begingroup
\sloppy
\bibliographystyle{elsarticle-num}
\bibliography{references}
\endgroup

\end{document}